\documentstyle[11pt,amsfonts]{article}

\newcommand{\be}{\begin{equation}}
\newcommand{\ee}{\end{equation}}
\newcommand{\bea}{\begin{array}}
\newcommand{\ea}{\end{array}}
\newcommand{\beqa}{\begin{eqnarray}}
\newcommand{\eeqa}{\end{eqnarray}}
\newcommand{\bean}{\begin{eqnarray*}}
\newcommand{\eean}{\end{eqnarray*}}

\def\up#1{\leavevmode \raise.16ex\hbox{#1}}

\def\sqr#1#2{{\vcenter{\vbox{\hrule height.#2pt
        \hbox{\vrule width.#2pt height#1pt \kern#1pt
          \vrule width.#2pt}
        \hrule height.#2pt}}}}

\newcommand{\gapproxeq}{\lower .7ex\hbox{$\;\stackrel{\textstyle
>}{\sim}\;$}}
\newcommand{\lapproxeq}{\lower .7ex\hbox{$\;\stackrel{\textstyle
<}{\sim}\;$}}

\newcounter{appendice}

\def\thebibliography#1{{\bf REFERENCES\markboth
 {REFERENCES}{REFERENCES}}\list
 {[\arabic{enumi}]}{\settowidth\labelwidth{[#1]}\leftmargin\labelwidth
 \advance\leftmargin\labelsep
 \usecounter{enumi}}
 \def\newblock{\hskip .11em plus .33em minus -.07em}
 \sloppy
 \sfcode`\.=1000\relax}

\begin{document}


\centerline{ \LARGE  Fate of the Born-Infeld solution in string theory} 

\vskip 2cm

\centerline{ {\sc G. Karatheodoris,  A. Pinzul and A. Stern}  }

\vskip 1cm
\begin{center}
Department of Physics, University of Alabama,\\
Tuscaloosa, Alabama 35487, USA
\end{center}

\vskip 2cm

\vspace*{5mm}

\normalsize
\centerline{\bf ABSTRACT} 
 We argue that the
 Born-Infeld solution   on the   D$9-$brane is  unstable under inclusion of 
 derivative corrections to  Born-Infeld theory   coming from
string theory.  More specifically, we find no electrostatic solutions
to the first  order corrected Born-Infeld theory on the  D$9-$brane which give
 a finite value for the Lagrangian.

\vspace*{5mm}

\newpage
\scrollmode

The Born-Infeld (BI) nonlinear description of electrodynamics is of current
interest due to its role as an effective action for
Dp-branes.\cite{Fradkin:1985},\cite{Johnson}  One of its main features
is the existence of a maximum allowed value $|E|_{max}$
for electrostatic fields.\footnote{Here  we don't consider the
  excitation of  transverse modes of the brane.}
  In the $30's$,
  Born and Infeld promoted this theory because it has a 
  spherically symmetric
solution with finite
classical self-energy.\cite{bi}  It agrees with the Coulumb solution
at large distances, but 
has a finite limit for the radial component of the electric field at
the origin.  This limit is just $|E|_{max}$.   As a result, the vector
field  is singular  at  the origin.  Equivalently, the solution
 is thus defined on a Euclidean manifold with one point (corresponding to the `source') removed,
 despite
 the fact that the  energy density is  well behaved at the point.  The
 theory contains one dimensionful parameter (namely,  $|E|_{max}$), which was 
determined by Born  and Infeld after fixing the classical self-energy
with the electron mass.  For the Dp-brane the
dimensionful parameter is the string tension, and so any BI-type
solution (or BIon\cite{Gib}) appearing there should 
correspond to a  charged object with characteristic mass at the
string tension.  The fact that energetics admits vector fields that are not everywhere defined  indicates that  BI theory cannot be a complete
description.\cite{inc}  Moreover, in string theory, the BI effective action is only valid for slowly varying
fields, i.e. it is the lowest order term in the derivative expansion
for the full effective  Dp-brane action.
Since derivatives of the  fields are not `small' in the interior of the BI
solution, the validity of such a solution in string theory can be questioned.
It is therefore of interest to know whether or not analogues of the
BI solutions survive for the full effective Dp-brane action.  A derivative  expansion
has been recently carried out  to obtain lowest order  string corrections
to the BI action for the space-filling D$9$-brane.\cite{Wyl},\cite{das}
With the restriction to electrostatic configurations, we find that the
BI solution on the D$9$-brane is unstable with the inclusion
of such corrections, indicating   that such
singular field configurations may not follow from the full effective
 action.  The reason is basically due to the result that
derivative corrections make it
 difficult for the electrostatic field to attain its
maximum value  $|E|_{max}$.  Here we only allow for field
configurations that lead to a
finite value for the Lagrangian.\footnote{We are unable to make a
  similar requirement for the energy, due to the fact
that the corrections to the action contain terms with multiple time
derivatives, making it  problematic to find the corresponding
Hamiltonian.}  This is a reasonable requirement from the point of view
of the path integral, where one expects to recover the classical
solutions in the WKB approximation.
 Moreover, we find no nontrivial electrostatic solutions
on the D$9$-brane D$9$-brane associated with a finite Lagrangian.  

The situation here is in contrast to skyrmion physics, where there are no nontrivial solutions to the zeroth
order effective action for QCD.  Higher order derivative
corrections, like the Skyrme term, are necessary to stabilize the
skyrmion.  On the other hand, BIons appear at lowest order, but
become unstable upon including the next order electrostatic corrections.
There remains the possibility, however, of stabilizing the BIon with the inclusion of other
degrees of freedom.  For example, if we allow for magnetic effects one
might find that the higher order corrections  give a magnetic
dipole moment to the BIon.  Another possibility which is of current
interest, concerns BI-type
solutions that
appear after dimensional reduction.  In this case, the BI action is replaced
by the Dirac-Born-Infeld action (DBI), containing   degrees
of freedom associated with the transverse modes of the brane.
Classical solutions to the DBI action  were found in  \cite{Gib},\cite{Callan:1997kz},
\cite{Howe:1997ue}, and they represent fundamental strings attached to
branes.\footnote{As no such interpretation is possible for the BIon on
  the D$9$-brane, it is convenient that we find  it to be unstable.}   It is of interest to examine how
such solutions are affected by  derivative corrections\cite{Wyl}.  (One set
of solutions (the BPS solutions) were found to be unaffected to all orders\cite{Thorlacius:1997zd}.)  We hope to address these issues in
future works.

We begin with a review of the Born-Infeld electrostatics. The BI action is expressed in terms of the determinant of the
matrix with elements
\be h_{\mu\nu} =\eta_{\mu\nu} + (2\pi \alpha') F_{\mu\nu}\;, \ee
 where $
F_{\mu\nu}=\partial_\mu{\cal A}_\nu-\partial_\nu{\cal A}_\mu $ is the field strength and one assumes the
flat metric $\eta_{\mu\nu}$.
On the D$9-$brane it  is given by
\be {\cal S}^{(0)}_{BI} = \frac1{(4\pi^2 \alpha ')^5 g_s} \int d^{d+1}x\; {\cal L}^{(0)}_{BI} \;,\qquad {\cal L}^{(0)}_{BI} =1 -
\sqrt{-\det [h_{\mu\nu}]}\;, \label{tbia} \ee where $d$  is $9$.
In the absence of magnetic fields ${\cal L}^{(0)}_{BI}$ simplifies to
\be {\cal L}^{(0)}_{BI} =1 -
\sqrt{1-{\vec f}^2} \;, \ee where $\vec f =2\pi\alpha' \vec E$, and
$\vec E$ is the electric field strength.   It is only defined for $ |\vec E|$
below  $|E|_{max}=(2\pi\alpha')^{-1}$.  Field strengths above
this
critical value are said to be associated with string instabilities.
For  nine dimensional spherically symmetric electrostatic configurations, we
define the angular variable $\theta$, taking values between $\pm
\pi/2$. It is a function of the radial
coordinate, with $\vec f =\hat r  \sin \theta (r) $.  Then  the
Lagrangian can be written
\be L_{BI}^{(0)} = \int d^dx \; {\cal L}^{(0)}_{BI}  = \Omega^{d-1} \int dr  r^{d-1}
\biggl(1-\cos\theta(r)\biggr)  \;,\label{efbi} \ee  where $\Omega^{d-1} $ is the volume of a unit $d-1$-sphere.
The relevant degree of freedom  is the potential ${\cal A}_0(r)$, where $ \sin
\theta(r)= {\cal A}'_0(r)$,  the prime   denoting a derivative in $r$.  The action is extremized with respect to ${\cal A}_0$ by
 \be \theta (r) = \tan^{-1}{\frac Q{r^{d-1}}}\;.\label{sfbi} \ee 
The integration constant $Q$ is the charge. (\ref{sfbi})
 approaches the  Coulomb solution when $r\rightarrow
\infty$.    When $r \rightarrow 0$, $\theta \rightarrow \frac\pi 2\;
{\rm sign}(Q)$, and so unlike in Maxwell theory, the Lagrangian
density has a finite value at the location of the source.

Concerning the energy, one can apply the canonical formalism starting
  from the action (\ref{tbia}).   The Hamiltonian density is 
\beqa {\cal H}^{(0)}_{BI} &=& \Pi^i \partial_0 {\cal A}_i - {\cal
  L}^{(0)}_{BI}\cr & &\cr &=& \Pi^i F_{0i} +\sqrt{-\det [h_{\mu\nu}]}-1 +  \Pi^i \partial_i {\cal A}_0  \label{ham}\;,\eeqa
  where $i=1,2,...,d$,
\be \Pi^i = -\pi\alpha' \sqrt{-\det [h_{\mu\nu}]}\; (h^{i0}-h^{0i})
  \;,\label{mca}\ee   are the  momenta conjugate to ${\cal
  A}_i$ and $h^{\mu\nu} h_{\nu \rho} = \delta^\mu_\rho$.  As usual, the momentum
 conjugate to ${\cal
  A}_0$ is constrained to be zero.  In the absence of magnetic fields,
(\ref{mca}) reduces to \be\vec \pi =\frac{ \vec\Pi}{2\pi \alpha'}
  =\frac {\vec f}{\sqrt{1-{\vec f}^2}}\;\;,\ee and so substituting into
  (\ref{ham}) gives \beqa {\cal H}^{(0)}_{BI}&=&
\sqrt{1+{\vec \pi}^2} -1\;  - \;(2\pi\alpha')  \partial_i\pi^i {\cal A}_0\cr & &\cr  &=&
\frac1{\sqrt{1-{\vec f}^2}} -1 \; -\;( 2\pi\alpha')  \partial_i\pi^i {\cal
  A}_0  \label{ham2}\;,\eeqa where we integrated by parts.  The
  coefficient of $ {\cal A}_0$ gives the Gauss law constraint.
 The remaining  terms can be used to identify the self-energy of the
 BI solution
\be E_{BI}^{(0)} =   \frac{\;\Omega^{d-1}}{(4\pi^2 \alpha ')^5 g_s}\int dr  r^{d-1}
\biggl(\sec\theta(r)-1 \biggr)  \;,\label{efbih} \ee
 which is finite, despite
the fact that the vector field $\vec f$   is ill-defined at the
origin.

  In addition to the  nine dimensional spherically symmetric solution,
  there are also  `axially symmetric' BI solutions.  They correspond
  to (\ref{sfbi}) with $d<9$.  Then instead of a point singularity on
  the $D9-$brane,
  there would be $(9-d)$ -dimensional Euclidean surface where  $\vec f$
  is singular.   In that case, (\ref{efbih}) is
the energy per unit volume along the surface, and $r$ is the distance from
the singular surface.

    A derivative  expansion
has been carried out  in \cite{Wyl},\cite{das} to obtain  corrections
to the BI action for the D$9$-brane.  At first order, one obtains terms involving first
and second derivatives of
$F_{\mu\nu}$.  They  are contained in the rank-$4$ tensor
\be  S_{\beta\gamma\nu\rho} =2\pi \alpha'\partial_\beta \partial_\gamma
F_{\nu\rho}
+(2\pi \alpha')^2  h^{\alpha \delta}( \partial_\beta F_{\nu \alpha} \partial_\gamma
F_{\rho\delta} -   \partial_\beta F_{\rho \alpha} \partial_\gamma
F_{\nu\delta} ) \;,\label{tens} \ee which is antisymmetric in the last two indices.
Up to  first order the  action is
$${\cal  S}^{(0)}_{BI} + {\cal S}^{(1)}_{BI} = \frac1{(4\pi^2 \alpha ')^5 g_s} \int
  d^{10}x\;\biggl\{1 -
\sqrt{-\det [h_{\mu\nu}]} \;\biggl(1 +\frac \kappa 4  \Delta\biggl)\;\biggr\} \;, $$
\be \Delta =\; h^{\mu\nu}  h^{\rho\sigma}h^{\alpha\beta}  h^{\gamma\delta}(
S_{\nu \rho\alpha\beta} S_{\sigma \mu\gamma\delta} - 2  S_{\beta\gamma\nu\rho}
S_{\delta\alpha\sigma\mu} )\;, \label{flfmfa}\ee where  $\kappa ={{
    (2\pi\alpha')^2}\over {48}} $.  The derivatives  $ \partial_\mu$ appearing in (\ref{tens}) must
be covariant derivatives with respect to diffeomorphism in the
ten dimensions  space-time for the action to be invariant under such
diffeos.\footnote{This fact was overlooked in  previous versions  of the paper.}
We avoid this complication and work in Cartesian coordinates $x_i$.
For  electrostatic fields  $\vec E(\vec x) =
\vec f(\vec x)/(2\pi\alpha')$,
\beqa
 S_{ijk0}=- S_{ij0k}&=&
\partial_i\partial_jf_k + \frac { f_\ell}{1-\vec f^2}
(\partial_if_k\partial_jf_\ell +\partial_if_\ell\partial_jf_k) \cr
 S_{ijk\ell}&=&
- \frac 1{1-\vec f^2} 
(\partial_if_k\partial_jf_\ell -\partial_if_\ell\partial_jf_k)\label{sijko}\eeqa
Then exploiting the symmetry of the indices we can write
\be -\frac14 \Delta =\frac12
S_{ijk\ell}S^{ijk\ell}+\frac1{1-\vec f^2} S_{ijk0}S_{\;\;\;\;\;0}^{ijk}\;, \ee where the spatial indices are raised
using $h^{ij} = \delta_{ij} + f_i f_j/(1- \vec f^2) $.

Next we again assume spherical symmetry $\vec f = \hat
r\sin\theta(r) $, $-\pi/2\le \theta \le \pi/2$.
After substituting into (\ref{sijko}) 
$$S_{ijk\ell}S^{ijk\ell}=\frac{2(d-1)}{r^4}\;\frac{\tan\theta}{\cos^3\theta}
\;[2(rH)^2 +
(d-2)\sin^2\theta]$$
$$ S_{ijk0}S_{\;\;\;\;\;0}^{ijk} =\frac 1{\cos^4\theta}\biggl[
H'^2+\frac{ d-1}{r^4}\biggl(
2 \; (rH -\sin\theta )^2+ (rH \cos^2\theta -\sin\theta)^2\biggr)
\biggl]\;,$$
where  $H=\theta'/\cos\theta$ and again $d=9$.
 Then the first order (in $\kappa$) corrected  Lagrangian is
given by \beqa  L_{BI}^{(0)}+ L_{BI}^{(1)} & =& \int d^{d-1}x \;\bigl( {\cal
  L}^{(0)}_{BI} + {\cal L}^{(0)}_{BI}\bigr) \cr  &  =& \Omega^{d-1}\int dr  r^{d-1}
\Biggl\{1-\cos\theta +\frac{\kappa  }{\cos^3\theta} \biggl(H'^2+
\frac{d-1}{r^4} \Sigma\biggr)\; \Biggr\}\;, \label{efbi2}\eeqa  where
\be \Sigma = 3(r H - \sin\theta)^2 +(rH \sin^2\theta)^2 + 2rH
\sin^3\theta + (d-2) \sin^4 \theta 
 \ee  The correction to the BI Lagrangian is positive (assuming
 $-\pi/2\le \theta \le \pi/2$), and so the appearance of the $\cos^3\theta$
 in the denominator makes it challenging to find solutions having a
 limiting value of $\pm \pi/2$ for $\theta$ and a finite value for $ L_{BI}^{(0)}+ L_{BI}^{(1)}$ .
The  action is extremized with respect to the potential ${\cal A}_0(r)$, where again $ \sin
\theta(r)= {\cal A}'_0(r)$, for 
$$  \frac1\kappa   ( Q - r^{d-1} \tan\theta)\;\cos^2\theta   =
 2\biggl(\frac{r^{d-1} H'}{\cos^3\theta}\biggr)'' + 3\;\frac{r^{d-1}
\sin\theta \;  H'^2}{\cos^3\theta}\qquad\qquad\qquad$$   $$ \qquad\qquad
- \frac{d-1}{4\cos^3\theta}\biggl\{ [27 - 4 \cos 2\theta + \cos
4\theta ](r^{d-3} H)' $$   
\be + \frac12 r^{d-5}\sin\theta  \biggl[126 -53 d + 85 r^2 H^2 
+ 4 \cos 2\theta (5 + d - 3 r^2 H^2)+   \cos 4\theta ( d - 2 - r^2 H^2)\biggr]  \biggr\}\label{fthord}\ee
We note that by rescaling $r$ and $Q$ we can set  $\kappa$  equal to unity.
 Also  (\ref{fthord}) is invariant
under $Q\rightarrow -Q$ and $\theta(r)\rightarrow -\theta(r)$, and
this gives a prescription for mapping any possible charged solution to
the anti-solution.
The left hand side of (\ref{fthord}) vanishes for the original BI  solution
(\ref{sfbi}), while the right hand side represents derivative
corrections.  Substituting (\ref{sfbi}) into the left hand side gives a vanishingly
small correction as $r\rightarrow\infty$, but it is singular for
$r\rightarrow 0$.  The BI solution  therefore cannot be
trusted near the origin.

Due to the presence of high order time derivatives in the correction
terms  in the action (\ref{flfmfa}), the  computation of the Hamiltonian
for the system, and consequently the correction to the electrostatic
energy (\ref{efbih}),
 is problematic.  For this reason  we only require
 solutions to  (\ref{fthord}) to be associated with a finite value
 for the Lagrangian  (\ref{efbi2}) instead  of the energy. 
Since the  BI
(actually, Coulumb) solution is valid  at large $r$,
 \be \theta (r)\rightarrow\frac Q{r^{d-1}}\;,\qquad{\rm as} \qquad
r\rightarrow\infty \;,\label{tbil} \ee we can use this as an initial
condition at some large value of $r$
and numerically  integrate to small $r$.  We find  that for $d=9$ and  $Q>0$,
$\theta(r)$  tends to
the limit  of $\pi/2$ as $r\rightarrow 0$, as with the BI solution.
However,  the corresponding radial Lagrangian density diverges as  $
 r\rightarrow 0$ (faster than $1/r$).
 We plot the results below for $Q=1$ (and $\kappa=1$):

\bigskip
\input epsf
\def\epsfsize#1#2{0.8#1}
\centerline{\hss{\epsffile{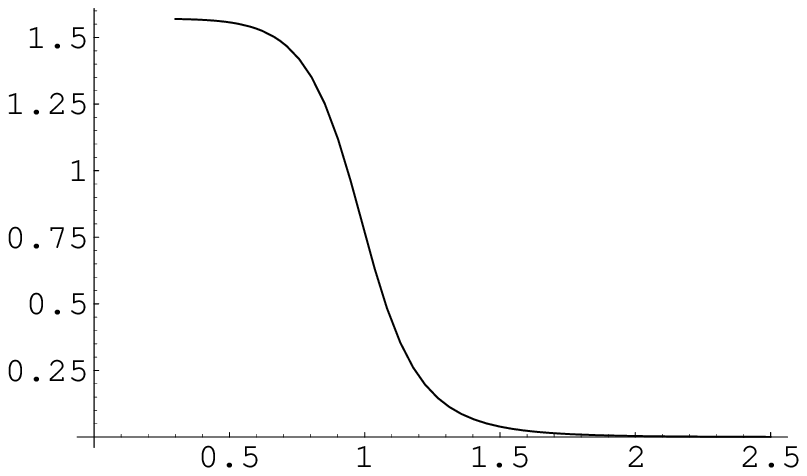}}\hss\hfill
  \hss{\epsffile{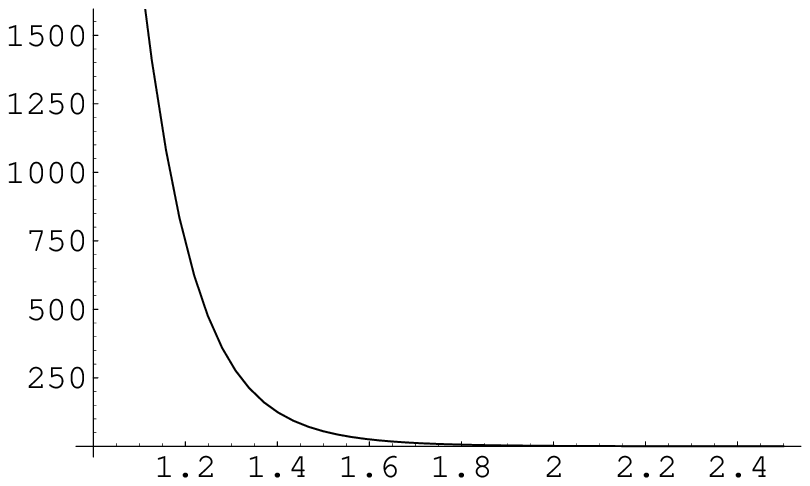}}\hss\hfill}

\medskip
\centerline{$\qquad\qquad {\tt fig. 1}\qquad\theta(r)\quad $ vs.$\quad r$
  \hfill $ {\tt fig. 2} \quad r^9\bigl( {\cal
  L}^{(0)}_{BI} + {\cal L}^{(0)}_{BI}\bigr)\;\; $ vs. $\;\;
r\qquad\qquad$}

\bigskip
\noindent
In Fig.2 we plot the radial Lagrangian density times $r$.  It blows up at
the origin, and so it appears that this solution does
 not lead to a finite value of the Lagrangian.  On the other hand,
the numerical integration procedure breaks down near the origin.  A
more
careful analysis requires that we look for a  solution near the
origin, consistent with the requirement of 
 a finite  Lagrangian, and try to match it with the above
 solution
for  some range of $r$.  The requirement of a finite Lagrangian
means that the limit of  $\theta(r) \rightarrow \pi /2$ as $r\rightarrow
0$ has to be approached slow enough so that the integral over the correction term
in  (\ref{efbi2}) is finite.  Upon writing $\theta(r) = \pi /2
-\epsilon (r)$, and keeping only the lowest order
terms in $\epsilon$, (\ref{fthord}) reduces to
\be
  (\epsilon Q - r^{d-1} )\;\epsilon^4 = 2\epsilon^3\biggl(\frac{r^{d-1} H'}{\epsilon^3}\biggr)'' + 3r^{d-1}
  H'^2
- (d-1)\biggl\{ 8  (r^{d-3} H)'  +  r^{d-5}  [13 -7 d + 12 r^2 H^2 
]  \biggr\}
 \;,
\label{eqnrz}\ee where $H' \rightarrow - (\log\epsilon)''$ and we again set
$\kappa$ equal to one.    Taking $d=9$ and $Q>0$, one has the following
 solution near the origin
\be \epsilon(r) \rightarrow \biggl(\frac{76432\;r^4}{125\; Q}\biggr)^{\frac15} \;,\qquad{\rm as} \qquad r\rightarrow 0 \label{sno}\ee
It is easy to check that it gives a
finite contribution to the Lagrangian as  $r\rightarrow 0$. (For this one only needs that
$\epsilon(r)$ goes to zero slower than $r^{5/3}$.)
 On the other hand, after numerically integrating
this solution starting from some initial value $r_0$ to increasing values of $r$ we find
 that for any value of $Q$, $\epsilon$ goes quickly to $\pi/2$ or $-\pi/2$,  and
 so the Lagrangian density is poorly behaved for large values of $r$.
 Below we plot the results for $Q=1$:

\bigskip
\input epsf
\def\epsfsize#1#2{0.8#1}
\centerline{\hss{\epsffile{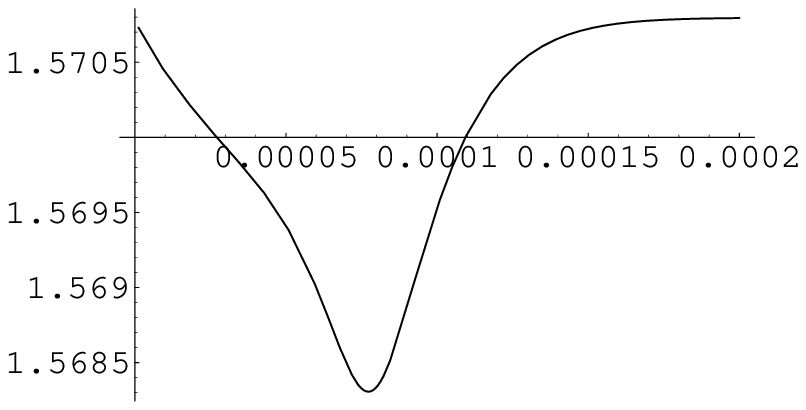}}\hss\hfill
  \hss{\epsffile{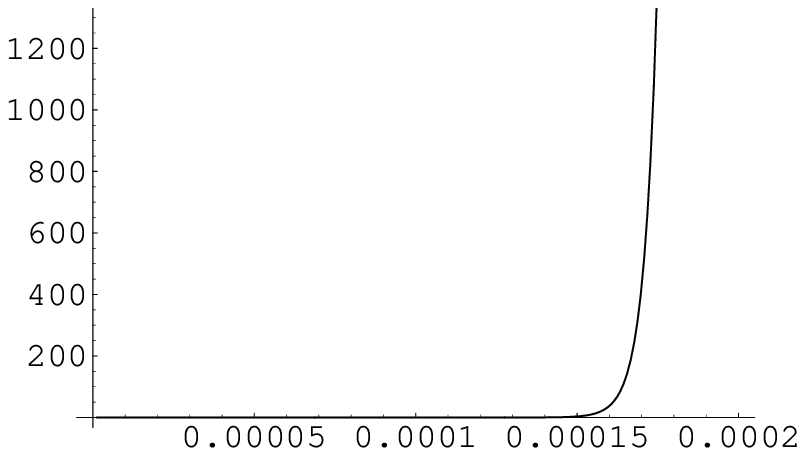}}\hss\hfill}

\medskip
\centerline{$\qquad \qquad{\tt fig. 3}\qquad\theta(r)\quad $ vs.$\quad r$
  \hfill $ {\tt fig. 4} \quad r^8\bigl( {\cal
  L}^{(0)}_{BI} + {\cal L}^{(0)}_{BI}\bigr)\;\; $ vs. $\;\;
r\qquad\qquad$}

\bigskip
\noindent
From fig. 3, $\theta'>0$ to the right of the minimum, and so no match with
fig. 1 is possible there.  To the left of the minimum, $\theta'$ tends
to
$-\infty$ as $r\rightarrow 0$ according to
 (\ref{sno}), while  from fig. 1 it appears to
vanish in the limit.   To make matters worse, the graphs we obtain
for $\theta$ near the origin are  highly sensitive to the initial
value $r_0$ of the integration, and this persists for all values of $Q$.  We have checked that this is not due to the neglect of
higher
order terms in (\ref{sno}).  So not only is there no agreement between the two numerical
integration procedures, the validity of the solution  near the origin
is questionable.
  We are thus unable to  find any
spherically symmetric
charged solutions to  (\ref{fthord}) consistent with the requirement of a  finite  Lagrangian.

 We also  find no  axially symmetric charged  solutions with finite Lagrangian
 density everywhere.  Again, they correspond
  to  $d<9$, and at zeroth order were characterized by a  $(9-d)$ -dimensional Euclidean surface where  $\vec f$
  is singular.   Using (\ref{tbil}) we can once again numerically integrate to
 small values of $r$.  The conclusions are the same as for $d=9$. For $Q>0$, $\theta(r)$  tends to
the limit  of $\pi/2$ as $r\rightarrow 0$, and  the corresponding radial Lagrangian density diverges as  $
 r\rightarrow 0$ (faster than $1/r$).
  For $Q>0$ and $d\ge 6$, the 
 solution to  (\ref{eqnrz}) near the origin has the form
\be \epsilon(r) \rightarrow \biggl(\frac{Cr^{d-5}}{125\; Q}\biggr)^{\frac15}
 \;,\qquad{\rm as} \qquad r\rightarrow 0\;, \label{snon}\ee where
 $C=20008,\;33324,\;51836$ for $d=6,7,8$, respectively.    As with the case $d=9$,
 it is consistent  with the requirement of a
finite Lagrangian for small $r$, but  the Lagrangian density  diverges after numerically integrating
 to increasing large $r$, and the problem of the sensitivity to the
 initial value $r_0$ persists. 
So again we find no agreement between the two numerical
integration procedures.  For  $d<6$, we find no  solutions
 to  (\ref{eqnrz}) near the origin satisfying the requirement of a
finite Lagrangian.

Above we found negative results for all  configurations associated
with some nonzero value for the charge.  Although there are no $Q=0$
soliton solutions of the zeroth order Born-Infeld system, this need
not be a priori  true at higher orders.    In this case, $\theta(r)$ for any
spherically symmetric solution would not fall
off as a power as $r\rightarrow \infty$, but rather
 \be \theta(r)\rightarrow \biggl\{A\cos{\alpha r} +B \sin{\alpha
   r}\biggl\}\frac{e^{-\alpha r}}{r^{(d-1)/4}}\;,\qquad{\rm as} \qquad
 r\rightarrow \infty\;,\label{noql}\ee
 where $\alpha=8^{-1/4}$, $A$ and $B$ are integration constants and we
 again set $\kappa = 1$.  Here we  look for  nonsingular soliton solutions,
and so   $\theta(r)$ should go to zero  as
$r\rightarrow 0$.  Such configurations then give a well
defined vector field at the origin.  Two types of behavior are allowed near the origin:
\be \theta(r)\rightarrow  C r\qquad{\rm or}\qquad  \theta(r)\rightarrow  D r^3\;,\qquad{\rm as} \qquad r\rightarrow 0\;,\label{noqs} \ee
where $C$ and $D$ are integration constants.   Next one can numerically integrate (\ref{fthord}) starting
from both
(\ref{noql}) and  (\ref{noqs}), and attempt to match
$\theta(r)$ and its derivatives for some range of $r$.  We found no
values for the constants $A$, $B$ and $C$ or $D$ where this is possible.

\bigskip

{\parindent 0cm{\bf Acknowledgement}}                                         

We are grateful for correspondence with Koji Hashimoto. 
This work was supported by the joint NSF-CONACyT grant E120.0462/2000 and
 the U.S. Department of Energy
 under contract number DE-FG05-84ER40141.


\begin{thebibliography}{99}

\bibitem{Fradkin:1985}
E.~S. Fradkin and A.~A. Tseytlin,  Phys. Lett. {\bf B163} (1985) 123; 
A.~Abouelsaood, C.~G. Callan, C.~R. Nappi, and S.~A. Yost, Nucl. Phys. {\bf B280} (1987) 599; 
E.~Bergshoeff, E.~Sezgin, C.~N. Pope, and P.~K. Townsend, Phys. Lett. {\bf
  188B} (1987) 70; 
R.~G. Leigh,  Mod.
  Phys. Lett. {\bf A4} (1989) 2767.
\bibitem{Johnson}  For reviews, see
C.~P.~Bachas,
arXiv:hep-th/9806199; A.~A.~Tseytlin,
arXiv:hep-th/9908105;
C.~V.~Johnson,
arXiv:hep-th/0007170; I.~V.~Vancea,
arXiv:hep-th/0109029; R.~J.~Szabo,
arXiv:hep-th/0207142.


\bibitem{bi} M. Born and L. Infeld, Proc. Roy. Soc. London {\bf A144} 425 (1934).

\bibitem{Gib}
G.~W.~Gibbons,
Nucl.\ Phys.\ B {\bf 514}, 603 (1998).



\bibitem{inc}
E. ~Feenberg, Phys.\ Rev. {\bf 47}, 148 (1935);
H.~L.~M. ~Pryce, Proc.\ Roy.\ Soc. {\bf A155}, 597 (1936);  
A.~A.~Chernitsky,
Helv.\ Phys.\ Acta {\bf 71}, 274 (1998);
D.~Chruscinski,
Phys.\ Lett.\ A {\bf 240}, 8 (1998).
\bibitem{Wyl}
N.~Wyllard,
Nucl.\ Phys.\ B {\bf 598}, 247 (2001); JHEP {\bf 0108}, 027 (2001).

\bibitem{das} S.~R.~Das, S.~Mukhi and N.~V.~Suryanarayana,
JHEP {\bf 0108}, 039 (2001).

\bibitem{Callan:1997kz}
C.~G.~Callan and J.~M.~Maldacena,
Nucl.\ Phys.\ B {\bf 513}, 198 (1998).

\bibitem{Howe:1997ue}
P.~S.~Howe, N.~D.~Lambert and P.~C.~West,
Nucl.\ Phys.\ B {\bf 515}, 203 (1998).

\bibitem{Thorlacius:1997zd}
L.~Thorlacius,
Phys.\ Rev.\ Lett.\  {\bf 80}, 1588 (1998).

\end{thebibliography}
\end{document}